\documentclass[pageno]{jpaper}
\pdfoutput=1

\usepackage[normalem]{ulem}
\usepackage{listings}           
\usepackage{courier}            
\usepackage{datetime}
\usepackage{tikz}               
\usepackage[T1]{fontenc}

\lstdefinelanguage{Scala}{
  morestring=[b]",
  morestring=[b]',
  morecomment=[l]{//},
  morecomment=[s]{/*}{*/},
  morekeywords={val,var,map,mapRows,mapCols,zip,zipWithIndex,groupBy,sum,min,minBy,reduce,fold,for,flatMap,multiFold,groupByFold,filter,flatMap,do,while,slice,copy}
}

\lstdefinelanguage{PPL}{
  morestring=[b]",
  morestring=[b]',
  morecomment=[l]{//},
  morecomment=[s]{/*}{*/},
  morekeywords={mapRows,mapCols,zip,zipWithIndex,groupBy,sum,min,minBy,reduce,for,filter,do,while,slice,copy,multiFold,fold,groupByFold,map,flatMap}
}

\author{
  Raghu Prabhakar \and David Koeplinger \and Kevin Brown \and HyoukJoong Lee \\
  \and Christopher De Sa \and Christos Kozyrakis \and Kunle Olukotun \and \\
  \vspace{-0.35in}\\
  \small Pervasive Parallelism Laboratory \vspace{-0.02in}\\
  \small Stanford University\\
  \small \{raghup17, dkoeplin, kjbrown, hyouklee, cdesa, kozyraki, kunle\}@stanford.edu\\
}

\begin{document}

\title{Generating Configurable Hardware from Parallel Patterns }

\date{}
\maketitle

\thispagestyle{empty}

\begin{abstract}
In recent years the computing landscape has seen an increasing shift towards specialized accelerators. Field programmable gate arrays (FPGAs) are particularly promising as they offer significant performance and energy improvements compared to CPUs for a wide class of applications and are far more flexible than fixed-function ASICs. However, FPGAs are difficult to program. Traditional programming models for reconfigurable logic use low-level hardware description languages like Verilog and VHDL, which have none of the productivity features of modern software development languages but produce very efficient designs, and low-level software languages like C and OpenCL coupled with high-level synthesis (HLS) tools that typically produce designs that are far less efficient.

Functional languages with parallel patterns are a better fit for hardware generation because they both provide high-level abstractions
to programmers with little experience in hardware design and avoid many of the problems faced when generating hardware from imperative languages.  In this paper, we identify two optimizations that are important when using parallel patterns to generate hardware:
tiling and metapipelining.  We present a general representation of tiled parallel patterns, and provide rules for
automatically tiling patterns and generating metapipelines.
We demonstrate experimentally that these optimizations result in speedups up to
$40 \times$ on a set of benchmarks from the data analytics domain.

\end{abstract}

\section{Introduction}
\label{intro}

The slowdown of Moore's law and the end of Dennard scaling has forced a radical change in the
architectural landscape. Computing systems are becoming increasingly parallel and heterogeneous,
relying on larger numbers of cores and specialized accelerators. Field programmable gate arrays
(FPGAs) are particularly promising as an acceleration technology, as they can offer performance and energy improvements for a wide
class of applications while also providing the reprogrammability and flexibility of software. Applications which exhibit
large degrees of spatial and temporal locality and which contain relatively small amounts of control flow, such as those in the image processing~\cite{grull2014biomedical,bailey2011design}, financial analytics~\cite{mencer2011finding,de2015fpga,zhang2005reconfigurable},
and scientific computing domains~\cite{smith2005scientific,alam2007using,brown2007performance,zhuo2008high}, can especially benefit from hardware acceleration with FPGAs. FPGAs have also recently been used to accelerate personal assistant systems \cite{sirius} and machine learning algorithms like deep belief networks \cite{baidu, catapultdnn}.

The performance and energy advantages of FPGAs are now motivating the integration of reconfigurable logic into
data center computing infrastructures. Both Microsoft~\cite{catapult} and Baidu~\cite{baidu} have recently announced
such systems. These systems have initially been in the form of banks of FPGA accelerators which communicate with CPUs through Infiniband or
PCIe \cite{maxeler}. Work is also being done on heterogeneous motherboards with shared CPU-FPGA memory \cite{harp}. The recent acquisition of Altera by Intel suggests that systems with tighter, high performance on-chip integration of CPUs and FPGAs are now on the horizon.

\begin{figure*} \centering\includegraphics[clip=true,width=7in,trim=0in 0in
0in 0in]{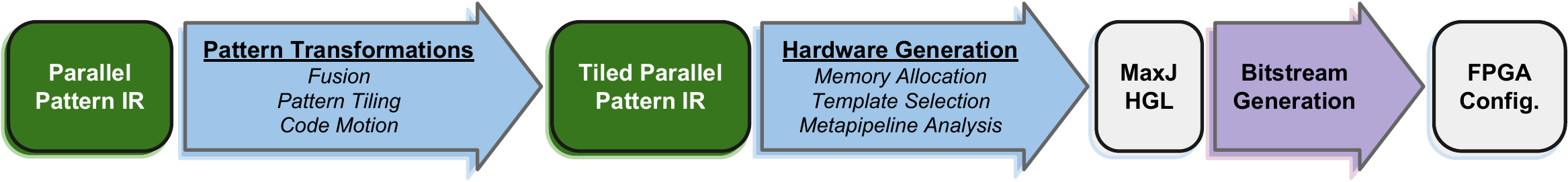}\caption{System diagram}
\label{fig:sysdiagram}
\end{figure*}

The chief limitation of using FPGAs as general purpose accelerators is that the programming model is currently inaccessible to most software developers.
Creating custom accelerator architectures on an FPGA is a complex task, requiring the coordination of large numbers of small,
local memories, communication with off-chip memory, and the synchronization of many compute stages.
Because of this complexity, attaining the best performance on FPGAs has traditionally required detailed hardware design using
hardware description languages (HDL) like Verilog and VHDL.
This low-level programming model has largely limited the creation of efficient custom hardware to experts in
digital logic and hardware design.

In the past ten years, FPGA vendors and researchers have attempted to make reconfigurable logic more accessible to software programmers with the
development of high-level synthesis (HLS) tools, designed to automatically infer register transaction level (RTL) specifications from higher level software programs. To better tailor these tools to software developers, HLS work has typically focused on imperative languages like C/C++, SystemC, and OpenCL~\cite{opencl}.
Unfortunately, there are numerous challenges in inferring hardware from imperative programs.
Imperative languages are inherently sequential and effectful. C programs in particular offer a number of challenges in alias analysis and detecting false dependencies~\cite{edwards}, typically requiring numerous user annotations to help HLS tools discover parallelism and determine when various hardware structures can be used.
Achieving efficient hardware with HLS tools often requires an iterative process to determine which user annotations are necessary, especially
for software developers less familiar with the intricacies of hardware design~\cite{cong11hls}.

Functional languages are a much more natural fit for high-level hardware generation as they have limited to no side effects and more naturally express a dataflow representation of applications which can be mapped directly to hardware pipelines \cite{fpgaMasses}. Furthermore, the order of operations in functional languages is only defined by data dependencies rather than sequential statement order, exposing significant fine-grained parallelism that can be exploited efficiently in custom hardware. 

Parallel patterns like \emph{map} and \emph{reduce} are an increasingly popular extension to functional languages which add semantic information about memory access
patterns and inherent data parallelism that is highly exploitable by both software and hardware.
Previous work~\cite{george14fpl, auerbach10lime} has shown that compilers can utilize parallel patterns to generate C- or OpenCL-based HLS programs and add certain annotations automatically.
However, just like hand-written HLS, the quality of the generated hardware is still highly variable.
Apart from the practical advantage of building on existing tools,
generating imperative code from a functional language only to have the HLS tool attempt to re-infer a
functional representation of the program is a suboptimal solution because higher-level semantic knowledge in
the original program is easily lost.
In this paper, we describe a series of compilation steps which automatically generate a low-level, efficient hardware design
from an intermediate representation (IR) based on parallel patterns.
As seen in Figure~\ref{fig:sysdiagram}, these steps fall into two categories: high level parallel pattern transformations (Section~\ref{transformations}),
and low level analyses and hardware generation optimizations (Section~\ref{hardware}).



One of the challenges in generating efficient hardware from high level programs is in handling arbitrarily large data structures. FPGAs have a
limited amount of fast local memory and accesses to main memory are expensive in terms of both performance and energy.
Loop tiling has been extensively studied as a solution to this problem, as it allows data structures with predictable access patterns to be broken up into fixed size
chunks. On FPGAs, these chunks can be stored locally in buffers. Tiling can also increase the reuse of these buffers by reordering computation,
thus reducing the number of total accesses to main memory.
Previous work on automated tiling transformations has focused almost exclusively
on imperative C-like programs with only affine, data-independent memory access patterns.
No unified procedure exists for automatically tiling a functional IR
with parallel patterns. In this paper, we outline a novel set of simple transformation
rules which can be used to automatically tile parallel patterns.
Because these rules rely on pattern matching rather than a mathematical model of the entire program, they can be used
even on programs which contain random and data-dependent accesses.


Our tiled intermediate representation exposes memory regions with high data locality, making them ideal candidates to be allocated on-chip. Parallel patterns provide
rich semantic information on the nature of the parallel computation at multiple levels of nesting as well as memory access patterns at each level.
In this work, we preserve certain semantic properties of memory regions and analyze memory access patterns in order to automatically infer
hardware structures like FIFOs, double buffers, and caches.
We exploit parallelism at multiple levels by automatically inferring and generating \emph{metapipelines}, hierarchical pipelines
where each stage can itself be composed of pipelines and other parallel constructs. Our code generation approach involves mapping parallel IR constructs
to a set of parameterizable hardware templates, where each template exploits a specific parallel pattern or memory access pattern.
These hardware templates are implemented using a low-level Java-based hardware generation language (HGL) called MaxJ.

In this paper we make the following contributions:
\begin{itemize}
  \item We describe a systematic set of rules for tiling parallel patterns, including a single, general pattern used to tile all patterns with fixed output size.
  Unlike previous automatic tiling work, these rules are based on pattern matching and therefore do not restrict all memory accesses within the program to be affine.

  \item We demonstrate a method for automatically inferring complex hardware structures like double buffers, caches, CAMs, and banked BRAMs from a parallel pattern IR.
  We also show how to automatically generate \emph{metapipelines}, which are a generalization of pipelines that greatly increase design throughput.

  \item
  We present experimental results for a set of benchmark applications from the data analytics domain running on an FPGA and show the performance impact of
  the transformations and hardware templates presented.
\end{itemize}


\section{Related Work}
\paragraph{Tiling}
Previous work on automated loop tiling has largely focused on tiling imperative programs
using polyhedral analysis~\cite{bondhugula08,pouchet10phd}.
There are many existing tools---such as Pluto~\cite{pluto08pldi},
PoCC~\cite{pouchet11popl}, CHiLL~\cite{chen2008chill},
and Polly~\cite{grosser2012polly}---that use polyhedral analysis
to automatically tile and parallelize programs.  These tools restrict memory
accesses within loops to only affine functions of the loop iterators.
As a consequence, while they perform well on affine sections of programs,
they fail on even simple, commonly occurring data-dependent operations
such as \emph{filters} and \emph{groupBys}~\cite{benabderrahmane10cc}. In order to handle these operations,
recent work has proposed using preprocessing steps which segment programs into affine
and non-affine sections prior to running polyhedral analysis tools~\cite{venkat}.

While the above work focused on the analysis of imperative programs, our work
analyzes functional parallel patterns, which offer a strictly higher-level representation
than simple imperative \emph{for} loops.
In this paper, we show that because of the additional semantic information
available in patterns like \emph{groupBy} and \emph{filter},
parallel patterns can be automatically tiled using
simple transformation rules, without the restriction that all memory accesses
are purely affine.
Little previous work has been done on automated tiling of functional
programs composed of arbitrarily nested parallel patterns.
Hielscher proposes a set of formal rules for tiling parallel operators \emph{map}, \emph{reduce}, and \emph{scan}
in the Parakeet JIT compiler, but these rules can be applied only for a small subset of nesting combinations \cite{parakeet}.
Spartan~\cite{spartan} is a runtime system with a set of high-level operators
(e.g., \emph{map} and \emph{reduce}) on multi-dimensional arrays, which
automatically tiles and distributes the arrays in a way that minimizes the
communication cost between nodes in cluster environments. In contrast to
our work, Spartan
focuses on distributed CPU computation, and not on optimizations that improve
performance on individual compute units.



\paragraph{Hardware from high-level languages}
Generating hardware from high-level languages has been widely studied for
decades.  CHiMPS~\cite{chimps} generates hardware from ANSI C code by
mapping each C language construct in a data-flow graph to an HDL block.
Kiwi~\cite{kiwi} translates a set of C\# parallel constructs (e.g.,
\emph{event}, \emph{monitor}, and \emph{lock}) to corresponding hardware units.
Bluespec~\cite{bluespec} generates hardware from purely functional descriptions
based on Haskell.  Chisel~\cite{chisel} is an embedded language in Scala
for hardware generation.  AutoPilot~\cite{autopilot} is a commercial HLS
tool that generates hardware from C/C++/SystemC languages.  Despite their
success in raising the level of abstraction compared to hardware description
languages, programmers are still required to write programs at a low-level and
express how computations are pipelined and/or parallelized.  Our work
abstracts away the implementation details from programmers by using high-level
parallel patterns, and applies compiler transformations and optimizations to
automatically pipeline and parallelize operations and exploit on-chip memory
for locality.

Recent work has explored using polyhedral analysis to optimize for data
locality on FPGAs~\cite{pouchet13fpga}.  Using polyhedral analysis, the compiler
is able to promote memory references to on-chip memory and parallelize
independent loop iterations with more hardware units.  However, the compiler is
not able to analyze loops that include non-affine accesses, limiting the
coverage of applications that can be generated for hardware. Our work can
handle parallel patterns with non-affine accesses by inferring required
hardware blocks (e.g., FIFOs and CAMs) for non-affine accesses, while
aggressively using on-chip memory for affine parts.

As high-level parallel patterns become increasingly popular to overcome the
shortcomings of C based languages, researchers have recently studied generating
hardware from functional parallel patterns.  Lime~\cite{auerbach10lime}
embeds high-level computational patterns (e.g., \emph{map}, \emph{reduce}, \emph{split}, and \emph{join}) in
Java and automatically targets CPUs, GPUs, and FPGAs without modifying the
code.  Our compiler manages a broader set of parallel patterns (e.g., \emph{groupBy})
and applies transformations even when patterns are nested,
which is common in a large number of real-world applications.  Recent work has
explored targeting nested parallel patterns to
FPGAs~\cite{george14fpl}. By exploiting the access patterns of nested patterns
to store sequential memory accesses to on-chip memory and parallelizing the
computation with strip-mining, the compiler can generate hardware that
efficiently utilizes memory bandwidth.  However, the compiler does not
automatically tile patterns for data locality or implement metapipelines
for nested parallel patterns, which we show are essential components to
generate efficient hardware. Our work is the first to show a method for
automatically tiling parallel patterns to improve locality and a process
for inferring hardware metapipelines from nested parallel patterns.



%
%

\section{Parallel Patterns}
\label{background}
\begin{figure*}
\small\centering
\begin{tabular}{l}
{\small\begin{tabular*}{0.95\textwidth}{ll|l}
\noalign{\hrule height 1.5pt}
{\bf Definitions} & & {\bf Usage Examples} \\ \hline

{\begin{lstlisting}[numbers=none,mathescape=true]
//Multidimensional
\end{lstlisting}} & & \\

{\begin{lstlisting}[numbers=none,mathescape=true]
Map(d)(m)
\end{lstlisting}} &

{\begin{lstlisting}[numbers=none,mathescape=true]
: V$_D$
\end{lstlisting}} &

{\begin{lstlisting}[numbers=none,mathescape=true]
x.map{ e => 2*e }; x.zip(y){ (a,b) => a + b }
\end{lstlisting}} \\

{\begin{lstlisting}[numbers=none,mathescape=true]
MultiFold(d)(r)(z)(f)(c)
\end{lstlisting}} &

{\begin{lstlisting}[numbers=none,mathescape=true]
: V$_R$
\end{lstlisting}} &

{\begin{lstlisting}[numbers=none,mathescape=true]
x.fold(1){ (a,b) => a * b }; x.map{ row => row.fold(0){ (a,b) => a + b } }
\end{lstlisting}} \\

{\begin{lstlisting}[numbers=none,mathescape=true]
//One-dimensional
\end{lstlisting}} & & \\

{\begin{lstlisting}[numbers=none,mathescape=true]
FlatMap(d)(n)
\end{lstlisting}} &

{\begin{lstlisting}[numbers=none,mathescape=true]
: V$_1$
\end{lstlisting}} &

{\begin{lstlisting}[numbers=none,mathescape=true]
x.flatMap{ e => if (e > 0) [e, -e] else [] }
\end{lstlisting}} \\

{\begin{lstlisting}[numbers=none,mathescape=true]
GroupByFold(d)(z)(g)(c)
\end{lstlisting}} &

{\begin{lstlisting}[numbers=none,mathescape=true]
: (K,V)$_1$
\end{lstlisting}} &

{\begin{lstlisting}[numbers=none,mathescape=true]
x.groupByFold(0){ e => (e/10, 1) }{ (a,b) => a + b }
\end{lstlisting}} \\ \hline

\end{tabular*}}
\\
\\
{\begin{tabular*}{0.95\textwidth}{llll}
{\bf User-defined Values} & & & \\ \hline
{\begin{lstlisting}[numbers=none,mathescape=true]
d : Integer$_D$
\end{lstlisting}} &

{\begin{lstlisting}[numbers=none,mathescape=true]
input domain
\end{lstlisting}} &

{\begin{lstlisting}[numbers=none,mathescape=true]
m : Index$_D$ => V
\end{lstlisting}} &

{\begin{lstlisting}[numbers=none,mathescape=true]
value function
\end{lstlisting}} \\

{\begin{lstlisting}[numbers=none,mathescape=true]
r : Integer$_R$
\end{lstlisting}} &

{\begin{lstlisting}[numbers=none,mathescape=true]
output range
\end{lstlisting}} &

{\begin{lstlisting}[numbers=none,mathescape=true]
n : Index => V$_1$
\end{lstlisting}} &

{\begin{lstlisting}[numbers=none,mathescape=true]
multi-value function
\end{lstlisting}} \\

{\begin{lstlisting}[numbers=none,mathescape=true]
z : V$_R$
\end{lstlisting}} &

{\begin{lstlisting}[numbers=none,mathescape=true]
init accumulator
\end{lstlisting}} &

{\begin{lstlisting}[numbers=none,mathescape=true]
f : Index$_D$ => (Index$_R$, V$_R$ => V$_R$)
\end{lstlisting}} &

{\begin{lstlisting}[numbers=none,mathescape=true]
(location, value) function
\end{lstlisting}} \\

{\begin{lstlisting}[numbers=none,mathescape=true]
c : (V$_R$,V$_R$) => V$_R$
\end{lstlisting}} &

{\begin{lstlisting}[numbers=none,mathescape=true]
combine accumulator
\end{lstlisting}} \hspace{51pt} &

{\begin{lstlisting}[numbers=none,mathescape=true]
g : Index => (K, V => V)$_1$
\end{lstlisting}} &

{\begin{lstlisting}[numbers=none,mathescape=true]
(key, value) function
\end{lstlisting}} \\
\noalign{\hrule height 1.5pt}

\end{tabular*}}
\\
\end{tabular}
\caption{\label{fig:ppl-syntax}Definitions and usage examples of supported parallel patterns.}
\end{figure*}
Parallel patterns are becoming a popular programming abstraction for writing high level applications that can still be efficiently mapped to hardware targets such as multicore~\cite{scala,haskell,delite-tecs14}, clusters~\cite{mapreduce,zaharia10spark,spartan}, GPUs~\cite{catanzaro11copperhead,micro14lee}, and FPGAs~\cite{auerbach10lime,george14fpl}. In addition, they have been shown to provide high productivity when implementing applications in a wide variety of domains~\cite{ecoop13sujeeth,pldi13halide}.
In this section we give a brief overview of the parallel patterns used in this paper.
We refer to the definitions presented in Figure~\ref{fig:ppl-syntax} as the parallel pattern language (PPL).
The definitions on the left represent the atoms in the intermediate language used in our compiler for analysis, optimization, and code generation. The code snippets on the right show common examples of how users typically interact with these patterns in a functional programming language via collections operations. The syntactic structure is essentially the same except that the input domain is inferred from the shape of the input collection.  Using explicit indices in the intermediate language allows us to model more user-facing patterns with fewer internal primitives
as well as express more complicated access patterns of the input data.

We separate our parallel patterns into two main groups. Multidimensional patterns have an arbitrary arity domain and range, but are restricted to have a range which is a fixed function of the domain. One-dimensional patterns on the other hand can have a dynamic output size. All patterns generate output values by applying a function to every index in the domain. Each pattern then merges these values into the final output in a different way. The output type $V$ can be a scalar or structure of scalars. We currently do not allow nested arrays, only multidimensional arrays. We denote multidimensional array types as $V_R$, which denotes a tensor of element type $V$ and arity $R$.  In Figure~\ref{fig:ppl-syntax} subscript $R$ always represents the arity of the output range, and $D$ the arity of the input domain.






\begin{figure}\centering
\begin{lstlisting}[language=Scala]
//data to be clustered, size n x d
val points: Array[Array[Float]] = ...

// current centroids, size k x d
val centroids: Array[Array[Float]] = ...

// Assign each point to the closest centroid by grouping
val groupedPoints = points.groupBy { pt1 =>
  // Assign current point to the closest centroid
  val minDistWithIndex = centroids.map { pt2 =>
    pt1.zip(pt2).map { case (a,b) => square(a - b) }.sum
  }.zipWithIndex.minBy(p => p._1)
  minDistWithIndex._2
}

// Average of points assigned to each centroid
val newCentroids = groupedPoints.map { case (k,v) =>
  v.reduce { (a,b) =>
    a.zip(b).map { case (x,y) => x + y }
  }.map { e => e / v.length }
}.toArray
\end{lstlisting}
\caption{$k$-means clustering implemented using Scala collections. In Scala, \texttt{\detokenize{_1}} and \texttt{\detokenize{_2}} refer to the first and second value contained within a tuple.}
\label{fig:kmeans}
\end{figure}

\begin{figure}\centering
\begin{lstlisting}
points: Array2D[Float](n,d) // data to be clustered
centroids: Array2D[Float](k,d) // current centroids

// Sum and number of points assigned to each centroid
(sums,counts) = multiFold(n)((k,d),k)(zeros((k,d),k)){ i =>
  pt = points.slice(i, *)
  // Assign current point to the closest centroid
  minDistWithIndex = fold(k)((max, -1)){ j =>
    pt2 = centroids.slice(j, *)
    dist = fold(d)(0){ p =>
      acc => acc + square(pt1(p) - pt2(p))
    }{ (a,b) => a + b }
    acc => if (acc._1 < dist) acc else (dist, j)
  }{ (a,b) => if (a._1 < b._1) a else b }

  minDistIndex = minDistWithIndex._2
  sumFunc = ((minDistIndex, 0), acc => {
    pt = points.slice(i, *)
    map(d){ j => acc(j) + pt(j) }
  })
  countFunc = (minDistIndex, acc => acc + 1)

  (sumFunc, countFunc)
}{ (a,b) => {
  pt = map(k,d){ (i,j) => a._1(i,j) + b._1(i,j) }
  count = map(k){ i => a._2(i) + b._2(i) }
  (pt, count)
} }

// Average assigned points to compute new centroids
newCentroids = map(k,d){ (i,j) =>
  sums(i,j) / counts(i)
}
\end{lstlisting}
\caption{$k$-means clustering using the parallel patterns in Figure~\ref{fig:ppl-syntax}. High level optimizations such as fusion have already been applied.}
\label{fig:kmeans-fused}
\end{figure}

\emph{Map} generates a single element per index, aggregating the results into a fixed-size output collection.
Note that the value function can close over an arbitrary number of input collections, and therefore this pattern is general enough to represent the classic operations \emph{map}, \emph{zip},
\emph{zipWithIndex}, etc.

\emph{MultiFold} is a generalization of a \emph{fold} which reduces generated values into a specified region of a (potentially) larger accumulator using an associative combine function.
The initial value $z$ is required to be an identity element of this function, and must have the same size and shape as the final output.
The main function $f$ generates an index specifying the location within the accumulator at which to reduce the generated value. We currently require the generated values to have the same arity as the full accumulator, but they may be of any size up to the size of the accumulator. Note that we can implement a traditional \emph{fold} as simply the special case where every generated value is the full size of the accumulator.
$f$ then converts each index into a function that consumes the specified slice of the current accumulator and returns the new slice. If the implementation maintains multiple partial accumulators in parallel, the combine function $c$ reduces them into the final result.

\begin{table*}[t]
\small\centering
\begin{tabular}{@{}lll@{}}
\toprule
{\bf Pattern }    & { }  & {\bf Strip Mined Pattern} \\ \midrule
{\begin{lstlisting}[mathescape=true,numbers=none,basicstyle=\fontsize{8}{8}\selectfont\tt]
T[ Map(d)(m) ]
\end{lstlisting}
} & \texttt{=} &
{\begin{lstlisting}[mathescape=true,numbers=none,basicstyle=\fontsize{8}{8}\selectfont\tt]
MultiFold(d/b)(d)(zeros(d)){ i =>
  (i, acc => Map(b)(T[m]) )
}(_)
\end{lstlisting}
} \\ \midrule
{\begin{lstlisting}[mathescape=true,numbers=none,basicstyle=\fontsize{8}{8}\selectfont\tt]
T[ MultiFold(d)(r)(z)(g)(c) ]
\end{lstlisting}
} & \texttt{=} &
{\begin{lstlisting}[mathescape=true,numbers=none,basicstyle=\fontsize{8}{8}\selectfont\tt]
MultiFold(d/b)(r)(T[z]){ i =>
  (i, acc => T[c](acc, MultiFold(b)(r)(T[z])(T[g])(T[c])) )
}(T[c])
\end{lstlisting}
} \\ \midrule
{\begin{lstlisting}[mathescape=true,numbers=none,basicstyle=\fontsize{8}{8}\selectfont\tt]
T[ GroupByFold(d)(z)(h)(c) ]
\end{lstlisting}
} & \texttt{=} &
{\begin{lstlisting}[mathescape=true,numbers=none,basicstyle=\fontsize{8}{8}\selectfont\tt]
GroupByFold(d/b)(T[z]){ i =>
  GroupByFold(b)(T[z])(T[h])(T[c])
}(T[c])
\end{lstlisting}
} \\ \midrule
{\begin{lstlisting}[mathescape=true,numbers=none,basicstyle=\fontsize{8}{8}\selectfont\tt]
T[ FlatMap(d)(f) ]
\end{lstlisting}
} & \texttt{=} &
{\begin{lstlisting}[mathescape=true,numbers=none,basicstyle=\fontsize{8}{8}\selectfont\tt]
FlatMap(d/b){i => FlatMap(b)(T[f]) }
\end{lstlisting}
} \\ \bottomrule
\end{tabular}
\caption{Strip mining transformation rules for parallel patterns defined in Figure \ref{fig:ppl-syntax}. }
\label{table:stripmining}
\end{table*}

\emph{FlatMap} is similar to \emph{Map} except that it can generate an arbitrary number of values per index.  These values are then all concatenated into a single flattened output. The output size can only be determined dynamically (it can be arbitrarily large) and therefore we restrict the operation to one-dimensional domains so that dynamically growing the output is easily defined. Note that this primitive also easily expresses a \emph{filter}.

\emph{GroupByFold} reduces generated values into one of many buckets where the bucket is selected by generating a key along with each value, i.e. it is a fused version of a \emph{groupBy} followed by a \emph{fold} over each bucket.  The operation is similar to \emph{MultiFold} except that the key-space cannot be determined in advance and so the output size is unknown. Therefore we also restrict this operation to one-dimensional domains.

Now that we have defined the operations, we will use them to implement $k$-means clustering as an example application.
For reference, first consider $k$-means implemented using the standard Scala collections operations, as shown in Figure~\ref{fig:kmeans}.
We will use this application as a running example throughout the remainder of this paper,
as it exemplifies many of the advantages of using parallel patterns as an abstraction
for generating efficient hardware.  $k$-means consumes a set of $n$ sample points of dimensionality $d$ and attempts to cluster those points by finding the $k$ best cluster centroids for the samples.  This is achieved by iteratively refining the centroid values. (We show only one iteration in Figure~\ref{fig:kmeans} for simplicity.)  First every sample point is assigned to the closest current centroid by computing the distance between every sample and every centroid. Then new centroid values are computed by averaging all the samples assigned to each centroid.  This process repeats until the centroid values stop changing. Previous work~\cite{rompf12optimizing,brown13clusters,chambers10flumejava} has shown how to automatically convert applications like $k$-means into a parallel pattern IR similar to ours as well as perform multiple high-level optimizations automatically on the IR.  One of the most important optimizations is fusing patterns together, both vertically (to decrease the reuse distance between producer-consumer relationships) and horizontally (to eliminate redundant traversals over the same domain).  Figure~\ref{fig:kmeans-fused} shows the structure of $k$-means after it has been lowered into PPL and fusion rules have been applied.  We have also converted the nested arrays in the Scala example to our multidimensional arrays.  This translation requires the insertion of \emph{slice} operations in certain locations, which produce a view of a subset of the underlying data.
For the remainder of this paper we will assume a high-level translation layer from user code to PPL exists and simply always start from the PPL description.

\section{Pattern Transformations}
\label{transformations}
One of the key challenges of generating efficient custom architectures from high level languages is in coping with arbitrarily large data structures. Since main memory accesses
are expensive and area is limited, our goal is to store a working set in the FPGA's local memory for as long as possible. Ideally, we also want
to hide memory transfer latencies by overlapping communication with computation using hardware blocks which automatically prefetch data.
To this end, in this section we describe a method for automatically tiling parallel patterns to improve program locality and data reuse.
Like classic loop tiling, our pattern tiling method is composed of two transformations: strip mining and interchange.
We assume here that our input is an intermediate representation
of a program in terms of optimized parallel patterns and that well known target-agnostic transformations like fusion, code motion, struct unwrapping, and common subexpression elimination
(CSE) have already been run.



\begin{table*}[t]
\small\centering
\begin{tabular}{@{}lll@{}}
\toprule
{\bf High Level Language}                            & {\bf PPL }       & {\bf Strip Mined PPL} \\ \midrule
{\begin{lstlisting}[numbers=none,language=Scala]
// Element-wise Map
val x: Array[Float] // length d
x.map{e => 2*e}
\end{lstlisting}}
&
{\begin{lstlisting}[numbers=none]
map(d){i => 2*x(i)}
\end{lstlisting}}
&
{\begin{lstlisting}[numbers=none]
multiFold(d/b)(d)(zeros(d)){ii =>
  xTile = x.copy(b + ii)
  (i, map(b)(b){i => 2*xTile(i) })
}(_)
\end{lstlisting}} \\ \midrule 
{\begin{lstlisting}[numbers=none, language=Scala]
// Sums along matrix rows
val x: Array[Array[Float]] // (m x n)
x.map{ row => 
  row.fold(0){ (a,b) => a + b } 
}
\end{lstlisting}}
&
{\begin{lstlisting}[numbers=none,language=PPL]
multiFold(m,n)(m)(zeros(m)){ (i,j) =>
  (i, acc => acc + x(i,j))
}{(a,b) => 
  map(n){(j) => a(j) + b(j)}
}
\end{lstlisting}}
&
{\begin{lstlisting}[numbers=none]
multiFold(m/b0,n/b1)(m)(zeros(m)){ (ii,jj) => 
  xTile = x.copy(b0 + ii, b1 + jj)
  tile = multiFold(b0,b1)(b0)(zeros(b0)){ (i,j) => 
    (i, acc => acc + xTile(i,j))
  }{(a,b) => map(b0){i => a(i) + b(i)} }
  (ii, acc => map(b0){j => acc(j) + tile(j)})
}{(a,b) =>
  multiFold(n/b0)(n)(zeros(n)){ii => 
    aTile = a.copy(b0 + ii)
    bTile = a.copy(b0 + ii)
    (ii, acc => map(b0){i => aTile(i) + bTile(i)})
  }{(a,b) => map(n){i => a(i) + b(i)}}
}
\end{lstlisting}} \\ \midrule 
{\begin{lstlisting}[numbers=none]
// Simple Filter
val x: Array[Float] // length d
x.flatMap{ e => 
  if (e > 0) e else [] 
}
\end{lstlisting}}
&
{\begin{lstlisting}[numbers=none]
flatMap(d){i => 
  if (x(i) > 0) x(i) else []
}
\end{lstlisting}}
&
{\begin{lstlisting}[numbers=none]
flatMap(d/b)(1){ii =>
  eTile = x.copy(b + ii)
  flatMap(b){i => 
    if (eTile(i) > 0) eTile(i) else [] 
}}
\end{lstlisting}} \\ \midrule 
{\begin{lstlisting}[numbers=none]
// Histogram Calculation
val x: Array[Float] // length d
x.groupByFold(0){ r => 
  (r/10, 1) 
}{ (a,b) => a + b }
\end{lstlisting}}
&
{\begin{lstlisting}[numbers=none]
groupByFold(d)(0){i => 
  (x(i)/10, 1)
}{(a,b) => a + b }
\end{lstlisting}}
&
{\begin{lstlisting}[numbers=none]
groupByFold(d/b)(0){ii => 
  xTile = x.copy(b + ii)
  groupByFold(b)(0){i => 
    (xTile(i)/10, 1)
  }{(a,b) => a + b}
}{(a,b) => a + b}
\end{lstlisting}} \\ \bottomrule
\end{tabular}
\caption{Examples of the parallel pattern strip mining transformation on Map, MultiFold, FlatMap, and GroupByFold}
\label{table:stripmine-examples}
\end{table*}

\paragraph{Strip mining} 

The strip mining algorithm is defined here using two passes over the IR.
The first strip mining pass partitions each pattern's iteration domain \emph{d} into tiles of size \emph{b} by breaking the pattern into a pair of perfectly nested patterns.
The outer pattern operates over the strided index domain, expressed here as \emph{d/b}, while the inner pattern operates on a tile of size \emph{b}.
For the sake of brevity this notation ignores the case where \emph{b} does not perfectly divide \emph{d}.
This case is trivially solved with the addition of \emph{min} checks on the domain of the inner loop.
Table~\ref{table:stripmining} gives an overview of the rules used to strip mine parallel patterns.
In addition to splitting up the domain, patterns are transformed by recursively strip mining all functions within that pattern.
Map is strip mined by reducing its domain and range and nesting it within a MultiFold. Note that the strided MultiFold writes
to each memory location only once. In this case we indicate the MultiFold's combination function as unused with an underscore.
As defined in Figure~\ref{fig:ppl-syntax}, the MultiFold, GroupByFold, and FlatMap patterns have the property that a perfectly nested form of a single instance of one of these
patterns is equivalent to a single ``flattened'' form of that same pattern. This property allows these patterns to be strip mined by
breaking them up into a set of perfectly nested patterns of the same type as the original pattern.

The second strip mining pass converts array slices and accesses with statically predictable access patterns into slices and accesses of larger, explicitly defined
array memory tiles. We define tiles which have a size statically known to fit on the FPGA using array copies.
Copies generated during strip mining can then be used to infer buffers during hardware generation.
Array tiles which have overlap, such as those generated from sliding windows in convolution, are marked with metadata in the IR as having some reuse factor.
Array copies with reuse have special generation rules to minimize the number of redundant reads to main memory when possible.

Table~\ref{table:stripmine-examples} demonstrates how our rules can be used to strip mine a set of simple data parallel operations.
We use the \emph{copy} infix function on arrays to designate array copies in these examples, using similar syntax as array \emph{slice}.
We assume in these examples that CSE and code motion transformation passes have been run after strip mining to eliminate duplicate copies and to
move array tiles out of the innermost patterns. In each of these examples, strip mining creates tiled copies of input collections that
we can later directly use to infer read buffers.

\paragraph{Pattern interchange}
Given an intermediate representation with strip mined nested parallel patterns, we now need to interchange patterns to increase the reuse
of newly created data tiles. This can be achieved by moving strided patterns out of unstrided patterns. However, as with imperative loops,
it is not sound to arbitrarily change the order of nested parallel patterns.
We use two rules for pattern interchange, adapted from the \emph{Collect}-\emph{Reduce} reordering rule in~\cite{brown13clusters}.
These rules both match on the special case of MultiFold where every iteration updates the entire accumulator, which we refer to here as a \emph{fold}.
The first interchange rule defines how to move a scalar, strided \emph{fold} out of an unstrided Map, transforming the nested loop into a strided \emph{fold} of a Map.
Note that this also changes the combination function of the \emph{fold} into a Map.
The second rule is the inverse of the first, allowing us to reorder a strided MultiFold with no reduction function (i.e. the outer pattern of a tiled Map)
out of an unstrided \emph{fold}. This creates a strided MultiFold of a scalar \emph{fold}. We apply these two rules whenever possible to increase the reuse
of tiled inputs.

Imperfectly nested parallel patterns commonly occur either due to the way the original user program was structured or
as a result of aggressive vertical fusion run prior to tiling.
Interchange on imperfectly nested patterns requires splitting patterns into perfectly nested sections. However, splitting and reordering trades
temporal locality of intermediate values for increased reuse of data tiles. In hardware, this can involve creating more main memory
reads or larger on-chip buffers for intermediate results so that less reads need to be done for input and output data. This tradeoff between memory reads and
increased buffer usage requires more complex cost modeling.
We use a simple heuristic to determine whether to split fused loops: we split and interchange patterns only when
the intermediate result created after splitting and interchanging is statically known to fit on the FPGA. This handles the simple case
where the FPGA has unused on-chip buffers and allocating more on-chip memory guarantees a decrease in the number of main memory reads.
Future work will examine ways to statically model the tradeoff between main memory accesses and local buffers when the chip is near 100\% on-chip memory utilization.

\begin{table*}[t]
\centering\small
\begin{tabular}{@{}lll@{}}
\toprule
{\bf High Level Language}                            & {\bf Strip Mined PPL }       & {\bf Interchanged PPL} \\ \midrule
{\begin{lstlisting}[numbers=none]
// Matrix Multiplication
x: Array[Array[Float]] // m x p
y: Array[Array[Float]] // p x n
x.map{row =>
  y.map{col =>
    row.zipWith(col){(a,b) =>
       a * b
    }.sum
  }
}
\end{lstlisting}}
&
{\begin{lstlisting}[numbers=none]
multiFold(m/b0,n/b1)(m,n)(zeros(m,n)){ (ii,jj) =>
  xTile = x.copy(b0 + ii, b1 + jj)
  ((ii,jj), acc =>
    map(b0,b1){ (i,j) =>
      multiFold(p/b2)(1)(0){ kk =>
        yTile = y.copy(b1 + jj, b2 + kk)
        accTile = fold(b2)(0){ k =>
          acc => acc + xTile(i,j) * yTile(j,k)
        }{(a,b) => a + b}
        (0, acc => acc + accTile)
        }{(a,b) => a + b})
      }{(a,b) => a + b}
    })
}(_)
\end{lstlisting}}
&
{\begin{lstlisting}[numbers=none]
multiFold(m/b0,n/b1)(m,n)(zeros(m,n)){(ii,jj) =>
  xTile = x.copy(b0 + ii, b1 + jj)
  ((ii,jj), acc =>
    multiFold(p/b2)(b0,b1)(zeros(b0,b1)){kk =>
      yTile = y.copy(b1 + jj, b2 + kk)
      (0, acc =>
        map(b0,b1){(i,j) =>
          acc(i,j) + fold(b2)(0){k =>
            acc => acc + xTile(i,j) * yTile(j,k)
        }{(a,b) => a + b}
      }
      (0, acc => map(b0,b1){(i,j) =>
                  acc(i,j) + tile(i,j) })
    }{(a,b) =>
      map(b0,b1){(i,j) => a(i,j) + b(i,j)}
    }
}(_)
\end{lstlisting}} \\ \bottomrule
\end{tabular}
\caption{Example of the pattern interchange transformation applied to matrix multiplication.}
\label{table:interchange-examples}
\end{table*}

Table~\ref{table:interchange-examples} shows a simple example of the application of our pattern interchange rules on matrix multiplication.
We assume here that code motion has been run again after pattern interchange has completed.
In matrix multiplication, we interchange the perfectly nested strided MultiFold and the unstrided Map.
This ordering increases the reuse of the copied tile of matrix \emph{y} and changes the scalar reduction into a tile-wise reduction.
Note that the partial result calculation and the inner reduction can now be vertically fused.

\paragraph{Discussion}
The rules we outline here for automatic tiling of parallel patterns are target-agnostic. However, tile copies should only be made explicit
for devices with scratchpad memory like FPGAs and GPUs. Architectures with hierarchical memory systems effectively maintain views of subsections of memory
automatically through caching, making explicit copies on these architectures a waste of both compute cycles and memory.

We currently require the user to explicitly specify tile sizes for all dimensions which require tiling. In future work, tile sizes for all pattern
dimensions will instead be determined by the compiler through automated tile size selection using modeling and design space exploration.

\begin{figure*}\small\centering
\hspace{-0.105\textwidth}
\centering\begin{tabular}{cm{0.45\textwidth}m{0.01\textwidth}m{0.44\textwidth}}
{} &
{\begin{lstlisting}
// Sum and number of points assigned to each centroid
(sums,counts) = multiFold(n/b0)((k,d),k)(...){ii =>
  pt1Tile = points.copy(b0 + ii, *)
  // Assign each point in tile to its closest centroid
  multiFold(b0)((k,d),k)(zeros(1,d),0){ i =>
    pt1 = pt1Tile.slice(i, *)
    // Assign current point to the closest centroid
    minDistWithIndex = multiFold(k/b1)(1)((max, -1)){ jj =>
      pt2Tile = centroids.copy(b1 + jj, *)
      // Find closest in centroid tile to current point
      minIndTile = fold(b1)((max,-1)){ j =>
        pt2 = pt2Tile.slice(j, *)
        dist = ... // Calculate distance between pt1 and pt2
        acc => if (acc._1 < dist) acc else (dist, j+jj)
      }{ (a,b) => if (a._1 < b._1) a else b }

      // Compare min dist in this tile to overall min
      (0, acc =>
        if (acc._1 < minIndTile._1) acc else minIndTile)
    }{(a,b) =>
      if (a._1 < b._1) a else b
    }


    minDistIndex = minDistWithIndex._2
    ... // Reduce pt1 into accumulator at minDistIndex
    ... // Increment count at minDistIndex
  }{(a,b) => ... // Tiled combination function }
  // Add sums and counts for this tile to accumulator
  (0, acc => ... // Tiled combination function
}{(a,b) => ... // Tiled combination function }

// Average assigned points to compute new centroids
newCentroids = multiFold(k/b1,d)(k,d)(...){ (ii,jj) =>
  sumsBlk = sums.copy(b1 + ii, *)
  countsBlk = counts.copy(b1 + ii)
  (ii, acc => map(k,d){ (i,j) =>
    sumsBlk(i,j) / countsBlk(i)
  })
}
\end{lstlisting}} & \hfill &
{\begin{lstlisting}
// Sum and number of points assigned to each centroid
(sums,counts) = multiFold(n/b0)((k,d),k)(...){ ii =>
  pt1Tile = points.copy(b0 + ii, *)
  // Assign each point in tile to its closest centroid
  minDistWithInds = multiFold(k/b1)(b1)(map(b1)((max, -1))){ jj =>
    pt2Tile = centroids.copy(b1 + jj, *)
    // Find closest in centroid tile to each point in point tile
    minIndsTile = map(b0){ i =>
      pt1 = pt1Tile.slice(i, *)
      // Find closest in centroid tile to current point
      minIndTile = fold(b1)((max,-1)){ j =>
        pt2 = pt2Tile.slice(j, *)
        dist = ... // Calculate distance between pt1 and pt2
        acc => if (acc._1 < dist) acc else (dist, j+jj)
      }{ (a,b) => if (a._1 < b._1) a else b }
    }
    // Each pt: compare min dist in centroid tile to overall min
    (0, acc => map(b0){ i =>
      if (acc(i)._1 < minIndsTile(i)._1) acc else minIndsTile(i) })
  }{(a,b) =>
    map(b0){i => if (a(i)._1 < b(i)._1) a(i) else b(i) }
  }
  // Sum and number for this tile of points
  multiFold(b0)(k,d)(zeros(k,d)){ i =>
    minDistIndex = minDistWithInds(i)._2
    ... // Reduce pt1 into accumulator at minDistIndex
    ... // Increment count at minDistIndex
  }{(a,b) => ... // Tiled combination function }
  // Add sums and counts for this tile to accumulator
  (0, acc => ... // Tiled combination function
}{(a,b) => ... // Tiled combination function }

// Average assigned points to compute new centroids
newCentroids = multiFold(k/b1,d)(k,d)(...){ (ii,jj) =>
  sumsBlk = sums.copy(b1 + ii, *)
  countsBlk = counts.copy(b1 + ii)
  (ii, acc => map(k,d){ (i,j) =>
    sumsBlk(i,j) / countsBlk(i)
  })
}
\end{lstlisting}}
\end{tabular}

\vspace{-0.2in}\begin{tabular}{cc}
{\parbox{0.45\textwidth}{\centering{(a) Strip mined $k$-means.}}} &
{\parbox{0.5\textwidth}{\centering{(b) Pattern Interchanged $k$-means.}}}
\end{tabular}

\vspace{0.1in}
\footnotesize{\begin{tabular}{|l|cc|cc|cc|}

\noalign{\hrule height 1.5pt}
\multicolumn{1}{|c|}{} & \multicolumn{2}{c|}{\bf Fused} & \multicolumn{2}{c|}{\bf Strip Mined}  & \multicolumn{2}{c|}{\bf Interchanged} \\
& Main Memory Reads & On-Chip Storage & Main Memory Reads & On-Chip Storage & Main Memory Reads & On-Chip Storage \\ \hline
\emph{points} & $n \times d$ & $d$ & $n \times d$ & $b_0 \times d$ & $n \times d$ & $b_0 \times d$ \\ \hline
\emph{centroids} & $n \times k \times d$ & $d$ & $n \times k \times d$ & $b_1 \times d$ & $(n/b_0) \times k \times d$ & $b_1 \times d$ \\ \hline
\emph{minDistWithIndex} & $0$ & $2$ & $0$ & $2$ & $0$ & $2 \times b_0$ \\
\noalign{\hrule height 1.5pt}
\end{tabular}}

\small
\vspace{0.1in}\centering{(c) Minimum number of words read from main memory and on-chip storage for data structures within $k$-means clustering after each IR transformation.}

\caption{Full tiling example for $k$-means clustering, starting from the fused representation in Figure \ref{fig:kmeans-fused}, using tile sizes of \emph{b$_0$} and \emph{b$_1$} for the number of points $n$ and the number of clusters $k$. The number of features $d$ is not tiled in this example.}
\label{fig:kmeans-example}
\end{figure*}

We conclude this section with a complete example of tiling the $k$-means clustering algorithm, starting from the fused representation shown in Figure~\ref{fig:kmeans-fused}. We assume here that we wish
to tile the number of input points, \emph{n}, with tile size \emph{b$_0$} and the number of clusters, \emph{k}, with tile size \emph{b$_1$} but not the number of dimensions, \emph{d}. This is representative of machine learning
classification problems where the number of input points and number of labels is large, but the number of features for each point is relatively small.

Figure~\ref{fig:kmeans-example} gives a comparison of the $k$-means clustering algorithm after strip mining and after pattern interchange. During strip mining, we create
tiles for both the \emph{points} and \emph{centroids} arrays, which helps us to take advantage of main memory burst reads. However, in the strip mined version, we still fully
calculate the closest centroid for each point. This requires the entirety of \emph{centroids} to be read for each point. We increase the reuse of each tile of \emph{centroids} by first splitting
the calculation of the closest centroid label from the MultiFold (Figure~\ref{fig:kmeans-example}a. line~5). The iteration over the centroids tile is then perfectly nested within the iteration over the points. Interchanging these two iterations allows us to reuse the centroids tile across points, thus decreasing the total number of main memory reads for this array by a factor of \emph{b$_0$}. This decrease comes
at the expense of changing the intermediate (distance, label) pair for a single point to a set of intermediate pairs for an entire tile of \emph{points}. Since the created intermediate result
has size 2\emph{b$_0$}, we statically determine that this is an advantageous tradeoff and use the split and interchanged form of the algorithm.

\section{Hardware Generation}
\label{hardware}
In this section, we describe how the tiled intermediate representation is translated into an efficient FPGA design.
FPGAs are composed of various logic, register, and memory resources.
These resources are typically configured for a specific hardware design using a hardware description language (HDL) that is translated into an FPGA configuration file.
Our approach to FPGA hardware generation translates our parallel pattern IR into MaxJ, a Java-based hardware generation language (HGL), which is in turn used to generate an HDL.
This is simpler than generating HDL directly because MaxJ performs tasks such as automatic pipelining and other low-level hardware optimizations.




Hardware generation follows a template-based approach. We analyze the structure of the parallel
patterns in the IR and determine the correct MaxJ template to translate the pattern
to hardware. Table~\ref{t-hwtemplates} lists the templates in three classes: memories, pipelined execution units,
and state machine controllers. We also summarize the parallel pattern IR construct whose behavior each template captures. Each template is intended
to capture a specific type of hardware execution functionality and can be composed with other templates. For example,
a \emph{Metapipeline controller} could be composed of multiple \emph{Parallel controller}s, each of which could
contain pipelined \emph{Vector} or \emph{Tree reduction} units.
We next describe the key features in the IR which we use to infer each of these template classes.

\begin{table*}[t]
\footnotesize\centering
\begin{tabular}{ |m{1.5cm}|l|m{8cm}|m{3.8cm}| }

\noalign{\hrule height 1.5pt}
\multicolumn{1}{|l}{} &
\multicolumn{1}{c}{\bf Template} &
\multicolumn{1}{c}{\bf Description} &
\multicolumn{1}{c|}{\bf IR Construct} \\
\noalign{\hrule height 1.0pt}

\multirow{3}{1.5cm}[0pt]{\centering \bf Memories}
& Buffer & On-chip scratchpad memory & Statically sized array \\ \cline{2-4}
& Double buffer & Buffer coupling two stages in a metapipeline & Same as metapipeline controller \\ \cline{2-4}
& Cache & Tagged memory to exploit locality in random memory access patterns & Non-affine accesses \\
\noalign{\hrule height 1.0pt}

\multirow{4}{1.5cm}[0pt]{\centering \bf Pipelined Execution Units}
& Vector & SIMD parallelism & Map over scalars \\ \cline{2-4}
& Reduction tree & Parallel reduction of associative operations & MultiFold over scalars \\ \cline{2-4}
& Parallel FIFO & Used to buffer ordered outputs of dynamic size & FlatMap over scalars \\ \cline{2-4}
& CAM & Fully associative key-value store & GroupByFold over scalars \\
\noalign{\hrule height 1.0pt}

\multirow{4}{1.5cm}[-10pt]{\centering \bf Controllers}
& Sequential & Controller which coordinates sequential execution & Sequential IR node \\ \cline{2-4}
& Parallel & Task parallel controller. Simultaneously starts all member modules when enabled, signals done when all members finish & Independent IR nodes \\ \cline{2-4}
& Metapipeline & Controller which coordinates execution of nested parallel patterns in a pipelined fashion & Outer parallel pattern with multiple inner patterns \\ \cline{2-4}
& Tile memory & Memory command generator to fetch tiles of data from off-chip memory & Transformer-inserted array copy \\
\noalign{\hrule height 1.5pt}

\end{tabular}
\caption{Hardware templates used in MaxJ code generation.}
\label{t-hwtemplates}
\end{table*}

\paragraph{Memory Allocation}
Generating efficient FPGA hardware requires effective usage of on-chip memories (buffers).
Prior to generating MaxJ, we run an analysis pass to allocate buffers for arrays based on data access patterns and size.
All arrays with statically known sizes, such as array \emph{copies} generated in the tiling transformation described in
Section~\ref{transformations}, are assigned to buffers. Dynamically sized arrays are kept in main memory and we generate
caches for any non-affine accesses to these arrays.
We also track the readers and writers of each memory and use this information
to instantiate a MaxJ template with the appropriate word width and number of ports.

\paragraph{Pipeline Execution Units}
We generate parallelized and pipelined hardware when parallel patterns compute with scalar values,
as occurs for the innermost patterns.
We implemented templates for each
pipelined execution unit in Table~\ref{t-hwtemplates} using MaxJ language
constructs, and instantiate each template with the proper parameters (e.g., data type,
vector length) associated with the parallel pattern.  The MaxJ compiler
applies low-level hardware optimizations such as vectorization, code
scheduling, and fine-grained pipelining, and generates efficient hardware.  For
example, we instantiate a reduction tree for a MultiFold over an
array of scalar values, which is automatically pipelined by the MaxJ compiler.

\paragraph{Metapipelining}
To generate high performance hardware from parallel patterns, it is insufficient to exploit only a single degree of parallelism.
However, exploiting nested parallelism requires mechanisms to orchestrate
the flow of data through multiple pipeline stages while also exploiting parallelism at each stage of execution,
necessitating a hierarchy of pipelines, or \emph{metapipeline}.
This is in contrast to traditional HLS tools which require inner patterns to have a static size and be completely unrolled in order to generate a flat pipeline containing both the inner and outer patterns.




We create metapipeline schedules by first performing a topological sort on the IR of the body of the current parallel pattern.
The result is a list of stages, where each stage contains a list of patterns which can be run concurrently.
Exploiting the pattern's semantic information, we then
optimize the metapipeline schedule by removing unnecessary memory transfers and redundant computations.
For instance, if the output memory region of the pattern has been assigned to a buffer,
we do not generate unnecessary writes to main memory.

As another example, our functional representation of tiled parallel patterns can sometimes create redundant accumulation functions,
e.g., in cases where a MultiFold is tiled into a nested MultiFold. During scheduling we identify
this redundancy and emit a single copy of the accumulator, removing the unnecessary intermediate buffer.
Finally, in cases where the accumulator of a MultiFold cannot completely fit on-chip, we add a special
forwarding path between the stages containing the accumulator. This optimization avoids redundant writes to memory and
reuses the current tile.
Once we have a final schedule for the metapipeline, we promote every output buffer in each stage
to a double buffer, which is required to avoid write after read (WAR) hazards between metapipeline stages.

\begin{figure*} \centering\includegraphics[clip=true,width=7in,trim=0in 0in
0in 0in]{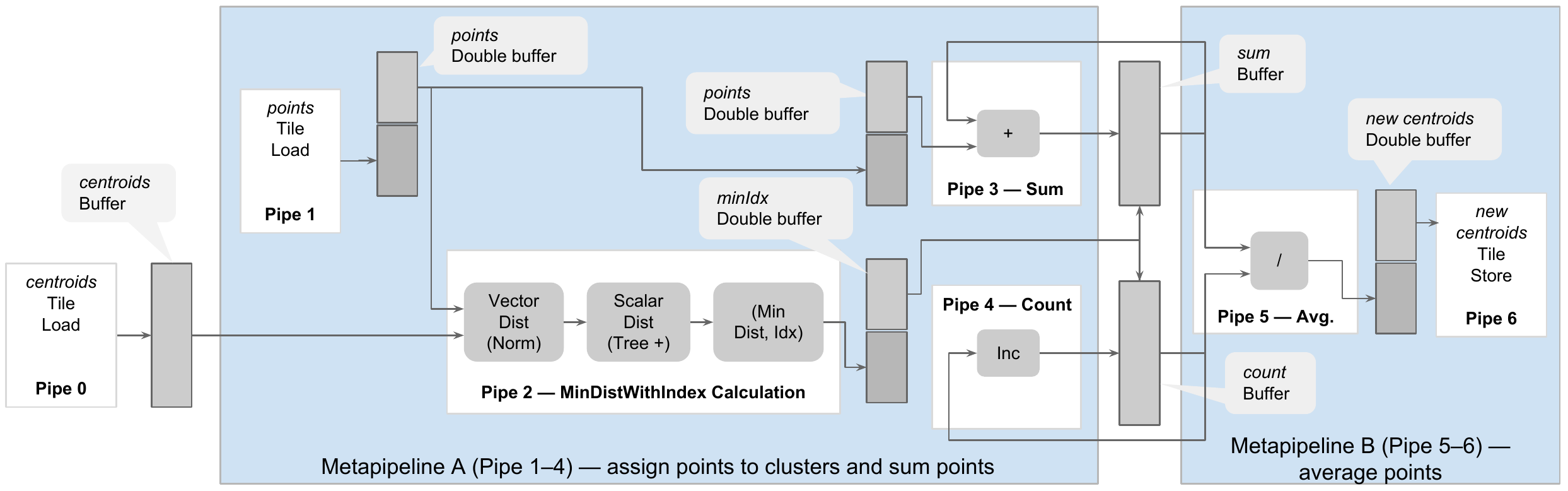}\caption{Hardware generated for the $k$-means application.}
\label{fig:metapipelining}
\end{figure*}

\paragraph{Example}
Figure~\ref{fig:metapipelining} shows a block diagram of the hardware generated for the $k$-means application.
For simplicity, this diagram shows the case where the \emph{centroids} array completely fits on-chip, meaning
we do not tile either the number of clusters \emph{k} or the number of features \emph{d}.
The generated hardware contains three sequential steps. The first step (Pipe~0) preloads the entire \emph{centroids} array into a buffer.
The second step (Metapipeline A) is a metapipeline which consists of three stages with double buffers to manage communication between the stages.
These three stages directly correspond to the three main sections of the MultiFold (Figure~\ref{fig:kmeans-fused}, line~5) used to sum and count the input points as grouped by their
closest centroid. The first stage (Pipe~1) loads a tile of the \emph{points} array onto the FPGA. Note that this stage is double buffered to
enable hardware prefetching. The second stage (Pipe~2) computes the index of the closest centroid using vector compute blocks and a scalar reduction
tree. The third stage (Pipe~3 and Pipe~4) increments the count for this minimum index and adds the current point to the corresponding location in the
buffer allocated for the \emph{new centroids}.
The third step (Metapipeline B) corresponds with the second outermost parallel pattern in the $k$-means application.
This step streams through the point sums and the centroid counts, dividing each sum by its corresponding count. The resulting new centroids
are then written back to main memory using a tile store unit for further use on the CPU.

Our automatically generated hardware design for the core computation of $k$-means is extremely similar to the manually optimized design described by Hussain et al.~\cite{hwkmeans}.
While the manual implementation assumes a fixed number of clusters and a small input dataset which can be preloaded onto the FPGA, we use tiling to automatically generate
buffers and tile load units to handle arbitrarily sized data. Like the manual implementation, we automatically parallelize across centroids
and vectorize the point distance calculations. As we see from the $k$-means example, our approach enables us to automatically generate high quality hardware implementations which are comparable to manual designs.



\section{Evaluation}
\label{evaluation}

\definecolor{bar1}{HTML}{5381BB}
\definecolor{bar2}{HTML}{BF4D4D}
\definecolor{bar3}{HTML}{9ABC5F}
\definecolor{bar4}{HTML}{8163A0}

\begin{figure*}[ht]
\centering

\includegraphics[width=\textwidth]{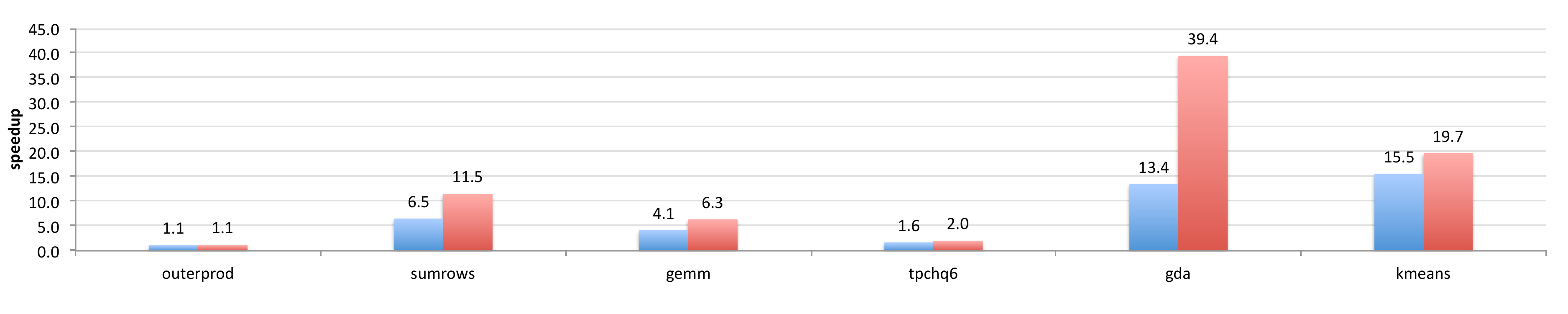}

\vspace{-10pt}

\includegraphics[width=\textwidth]{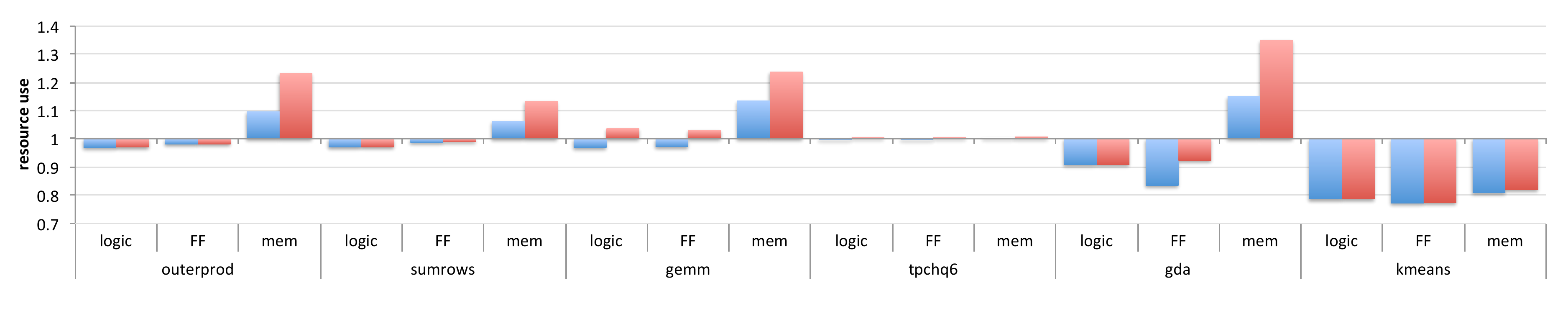}

{
\fontfamily{phv}\selectfont
\footnotesize
\raisebox{-0.2em}{\tikz{\path[fill=bar1] (0,0) rectangle (1em,1em);}}
+tiling
\hspace{2em}
\raisebox{-0.2em}{\tikz{\path[fill=bar2] (0,0) rectangle (1em,1em);}}
+tiling+metapipelining
}

\caption{Speedups and relative resource usages, relative to base design, resulting from our optimizations.}
\label{fig:speedup-bars}
\end{figure*}

We evaluate our approach to hardware generation described in Sections~\ref{transformations} and \ref{hardware} by comparing the performance and
area utilization of the FPGA implementations of a set of data analytic benchmarks.
We focus our investigation on the relative improvements that tiling and metapipelining provide over hardware designs that do not have these features.

\subsection{Methodology}
The benchmarks used in our evaluation are summarized in Table~\ref{table:benchmarks}.
We choose to study vector outer product, matrix row summation, and matrix multiplication as these exemplify many commonly occurring access patterns in the machine learning domain.
TPC-H Query 6 is a data querying application which reads a table of purchase records, filtering all
records which match a given predicate. It then computes the sum of a product of two columns in the filtered records.
Gaussian discriminant analysis (GDA) is a classification algorithm which models the distribution of each class as a multivariate Gaussian.
$k$-means clustering groups a set of input points by iteratively calculating the $k$ best cluster centroids.
In our implementations, all of these benchmarks operate on single precision, floating point input data.

We implement our transformation and hardware generation steps in an existing compiler framework called Delite \cite{delite-tecs14}.
We use the modified Delite compiler to generate MaxJ hardware designs as described in Sections \ref{transformations} and \ref{hardware}.
We then use the Maxeler MaxCompiler toolchain to generate an FPGA configuration bitstream from our generated MaxJ. We use the Maxeler runtime layer to manage communication with the FPGA from the host CPU.
We measure the running times of these designs starting after input data has been copied to the FPGA's DRAM and ending when the hardware design reports completion.
Final running times were calculated as an arithmetic mean of five individual run times.

We run each generated design on an Altera Stratix V FPGA on a Max4 Maia board.
The Maia board contains 48~GB of DDR3 DRAM with a maximum bandwidth of 76.8~GB/s.
The area numbers given in this section are obtained from synthesis reports provided by Altera's logic synthesis toolchain.
Area utilization is reported under three categories: Logic utilization (denoted ``logic''), flip flop usage (``FF''), and on-chip memory usage (``mem'').


\begin{table}
\centering\footnotesize
\begin{tabular}{lll}
\toprule

{\bf Benchmark} & {\bf Description} & {\bf Collections Ops} \\ \midrule
outerprod & Vector outer product & \emph{map} \\ \midrule
sumrows & Matrix summation through rows & \emph{map, reduce} \\ \midrule
gemm & Matrix multiplication & \emph{map, reduce} \\ \midrule
tpchq6 & TPC-H Query 6 & \emph{filter, reduce} \\ \midrule
gda & Gaussian discriminant analysis & \emph{map, filter, reduce} \\ \midrule
kmeans & $k$-means clustering & \emph{map, groupBy, reduce} \\ \bottomrule
\end{tabular}
\caption{Evaluation benchmarks with major collections operations used by
Scala implementation.}
\label{table:benchmarks}
\end{table}

\subsection{Experiments}
The baseline for each benchmark is an optimized hardware design implemented using MaxJ
that contains optimizations commonly used in state-of-the-art high-level synthesis
tools. In particular, each baseline design exploits data and pipelined parallelism within patterns where possible.
Pipelined parallelism is exploited for patterns that operate on scalars. Our baseline design
exploits locality at the level of a single DRAM burst, which on the MAX4 MAIA board is 384 bytes.
To isolate the effects of the amount of parallelism in our comparison, we keep
the innermost pattern parallelism factor constant between the baseline design and our optimized versions for each benchmark.

We evaluate our approach against the baseline by generating two hardware configurations per benchmark:
a configuration with tiling but no metapipelining, and a configuration with both tiling and metapipelining optimizations enabled.

\paragraph{Impact of tiling alone}
Figure \ref{fig:speedup-bars} shows the obtained speedups as well as relative on-chip resource utilizations for each of benchmarks.
As can be seen, most benchmarks in our suite show significant speedup when tiling
transformations are enabled. Benchmarks like \emph{sumrows} and \emph{gemm}
benefit from inherent locality in their memory accesses. For \emph{gemm}, our automatically generated code
achieves a speedup of $4\times$ speedup over the baseline for a marginal increase of about $10\%$ on-chip memory usage.


Benchmarks \emph{outerprod} and \emph{tpchq6} do not
show a significant difference with our tiling transformations over the baseline.
This is because both \emph{outerprod} and
\emph{tpchq6} are both memory-bound benchmarks. \emph{Tpchq6} streams through the input once without reuse, and streaming
data input is already exploited in our baseline design. Hence tiling does not provide any additional benefit.
The core compute pipeline in \emph{outerprod} is memory-bound at the stage writing results to DRAM, which cannot be addressed
using tiling. Despite the futility of tiling in terms of performance, tiling \emph{outerprod}
has a noticeable increase in memory utilization as the intermediate result varies as the square of the tile size.

In the cases of \emph{kmeans} and \emph{gda}, some
of the input data structures are small enough that they can be held in on-chip memory. This completely
eliminates accesses to off-chip memory, leading to dramatic speedups of $13.4\times$ and $15.5\times$ respectively
with our tiling transformations. \emph{gda} uses more on-chip memory to store intermediate data. On the other hand, the tiled
version \emph{kmeans} utilizes less on-chip memory resources. This is because the baseline for \emph{kmeans} instantiates multiple
load and store units, each of which creates several control structures in order to read and write data from DRAM. Each of these control
structures includes address and data streams, which require several on-chip buffers. By tiling, we require a smaller number of load and
store units.

\paragraph{Impact of metapipelining}
The second bar of Figure~\ref{fig:speedup-bars} shows the benefits of metapipelining. Metapipelines increase design throughput
at the expense of additional on-chip memory resources for double buffers.
Metapipelining overlaps the compute part of the design with the data transfer and hides the cost of the slower stage. Benchmarks like
\emph{gemm} and \emph{sumrows} naturally benefit from metapipelining because the memory transfer time is completely overlapped
with the compute, resulting in speedups of $6.3\times$ and $11.5\times$ respectively. Metapipelining also exploits overlap in
streaming benchmarks like \emph{tpchq6}, where the input data is fetched and stored simultaneously with the core computation.

Metapipelines with stages that are balanced, where each stage accounts for roughly equal number of cycles,
can benefit the most, as this achieves the most overlap. However, if the metapipeline stages are unbalanced, the throughput of the
pipeline is limited by the slowest stage, thus limiting performance improvement. We observe this behavior in \emph{outerprod},
where the metapipeline is bottlenecked at the stage writing results back to DRAM. On the other hand, applications
like \emph{gda}, \emph{kmeans} and \emph{sumrows} greatly benefit from metapipelining. In particular, \emph{gda} naturally
maps to nested metapipelines that are well-balanced. The stage loading the input tile overlaps execution with the stage
computing the output tile and the stage storing the output tile. The stage computing the output tile is also
a metapipeline where the stages perform vector subtraction, vector outer product and accumulation. We parallelize the vector
outer product stage as it is the most compute-heavy part of the algorithm; parallelizing the vector outer product enables
the metapipeline to achieve greater throughput. This yields an overall speedup of $39.4\times$
over the baseline.

\section{Conclusion}
In this paper, we introduced a set of compilation steps necessary
to produce an efficient FPGA hardware design from an intermediate representation composed of nested parallel patterns.
We described a set of simple transformation rules which can be used to automatically
tile parallel patterns which exploit semantic information inherent within these patterns
and which place fewer restrictions on the program's memory accesses than previous work.
We then presented a set of analysis and generation steps which can be used to automatically infer
optimized hardware designs with metapipelining.
Finally, we presented experimental results for a set of benchmarks in the machine learning and data querying
domains which show that these compilation steps provide performance improvements of up to $40 \times$ with a minimal impact on FPGA resource usage.

\bibliographystyle{plain}
\bibliography{references}

\begin{thebibliography}{10}

\bibitem{alam2007using}
Sadaf~R Alam, Pratul~K Agarwal, Melissa~C Smith, Jeffrey~S Vetter, and David
  Caliga.
\newblock Using fpga devices to accelerate biomolecular simulations.
\newblock {\em Computer}, (3):66--73, 2007.

\bibitem{bluespec}
Arvind.
\newblock Bluespec: A language for hardware design, simulation, synthesis and
  verification invited talk.
\newblock In {\em Proceedings of the First ACM and IEEE International
  Conference on Formal Methods and Models for Co-Design}, MEMOCODE '03, pages
  249--, Washington, DC, USA, 2003. IEEE Computer Society.

\bibitem{auerbach10lime}
Joshua Auerbach, David~F. Bacon, Perry Cheng, and Rodric Rabbah.
\newblock Lime: A java-compatible and synthesizable language for heterogeneous
  architectures.
\newblock In {\em Proceedings of the ACM International Conference on Object
  Oriented Programming Systems Languages and Applications}, OOPSLA '10, pages
  89--108, New York, NY, USA, 2010. ACM.

\bibitem{chisel}
J.~Bachrach, Huy Vo, B.~Richards, Yunsup Lee, A.~Waterman, R.~Avizienis,
  J.~Wawrzynek, and K.~Asanovic.
\newblock Chisel: Constructing hardware in a scala embedded language.
\newblock In {\em Design Automation Conference (DAC), 2012 49th ACM/EDAC/IEEE},
  pages 1212--1221, June 2012.

\bibitem{fpgaMasses}
David Bacon, Rodric Rabbah, and Sunil Shukla.
\newblock Fpga programming for the masses.
\newblock {\em Queue}, 11(2):40:40--40:52, February 2013.

\bibitem{bailey2011design}
Donald~G Bailey.
\newblock {\em Design for embedded image processing on FPGAs}.
\newblock John Wiley \& Sons, 2011.

\bibitem{benabderrahmane10cc}
Mohamed-Walid Benabderrahmane, Louis-No{\"e}l Pouchet, Albert Cohen, and
  C{\'e}dric Bastoul.
\newblock The polyhedral model is more widely applicable than you think.
\newblock In {\em ETAPS International Conference on Compiler Construction
  (CC'2010)}, pages 283--303, Paphos, Cyprus, March 2010. Springer Verlag.

\bibitem{bondhugula08}
Uday Bondhugula, Albert Hartono, J.~Ramanujam, and P.~Sadayappan.
\newblock A practical automatic polyhedral parallelizer and locality optimizer.
\newblock In {\em Proceedings of the 29th ACM SIGPLAN Conference on Programming
  Language Design and Implementation}, PLDI '08, pages 101--113, New York, NY,
  USA, 2008. ACM.

\bibitem{pluto08pldi}
Uday Bondhugula, Albert Hartono, J.~Ramanujam, and P.~Sadayappan.
\newblock A practical automatic polyhedral program optimization system.
\newblock In {\em ACM SIGPLAN Conference on Programming Language Design and
  Implementation (PLDI)}, June 2008.

\bibitem{brown13clusters}
Kevin~J. Brown, Arvind~K. Sujeeth, HyoukJoong Lee, Tiark Rompf, Christopher~De
  Sa, Martin Odersky, and Kunle Olukotun.
\newblock Big data analytics with delite.
\newblock \url{http://ppl.stanford.edu/papers/delite-scaladays13.pdf}, 2013.

\bibitem{brown2007performance}
Samuel Brown et~al.
\newblock Performance comparison of finite-difference modeling on cell, fpga
  and multi-core computers.
\newblock In {\em SEG/San Antonio Annual Meeting}, 2007.

\bibitem{catanzaro11copperhead}
Bryan Catanzaro, Michael Garland, and Kurt Keutzer.
\newblock Copperhead: compiling an embedded data parallel language.
\newblock In {\em Proceedings of the 16th ACM symposium on Principles and
  practice of parallel programming}, PPoPP, pages 47--56, New York, NY, USA,
  2011. ACM.

\bibitem{chambers10flumejava}
Craig Chambers, Ashish Raniwala, Frances Perry, Stephen Adams, Robert~R. Henry,
  Robert Bradshaw, and Nathan Weizenbaum.
\newblock Flumejava: easy, efficient data-parallel pipelines.
\newblock In {\em Proceedings of the 2010 ACM SIGPLAN conference on Programming
  language design and implementation}, PLDI. ACM, 2010.

\bibitem{chen2008chill}
Chun Chen, Jacqueline Chame, and Mary Hall.
\newblock Chill: A framework for composing high-level loop transformations.
\newblock Technical report, Citeseer, 2008.

\bibitem{cong11hls}
J.~Cong, Bin Liu, S.~Neuendorffer, J.~Noguera, K.~Vissers, and Zhiru Zhang.
\newblock High-level synthesis for fpgas: From prototyping to deployment.
\newblock {\em Computer-Aided Design of Integrated Circuits and Systems, IEEE
  Transactions on}, 30(4):473--491, April 2011.

\bibitem{de2015fpga}
Christian de~Schryver.
\newblock {\em FPGA Based Accelerators for Financial Applications}.
\newblock Springer, 2015.

\bibitem{mapreduce}
Jeffrey Dean and Sanjay Ghemawat.
\newblock {MapReduce: Simplified Data Processing on Large Clusters}.
\newblock In {\em OSDI}, OSDI, pages 137--150, 2004.

\bibitem{edwards}
S.A. Edwards.
\newblock The challenges of synthesizing hardware from c-like languages.
\newblock {\em Design Test of Computers, IEEE}, 23(5):375--386, May 2006.

\bibitem{george14fpl}
Nithin George, HyoukJoong Lee, David Novo, Tiark Rompf, Kevin~J. Brown,
  Arvind~K. Sujeeth, Martin Odersky, Kunle Olukotun, and Paolo Ienne.
\newblock Hardware system synthesis from domain-specific languages.
\newblock In {\em Field Programmable Logic and Applications (FPL), 2014 24th
  International Conference on}, pages 1--8, Sept 2014.

\bibitem{grosser2012polly}
Tobias Grosser, Armin Groesslinger, and Christian Lengauer.
\newblock Polly—performing polyhedral optimizations on a low-level
  intermediate representation.
\newblock {\em Parallel Processing Letters}, 22(04):1250010, 2012.

\bibitem{grull2014biomedical}
Frederik Grull and Udo Kebschull.
\newblock Biomedical image processing and reconstruction with dataflow
  computing on fpgas.
\newblock In {\em Field Programmable Logic and Applications (FPL), 2014 24th
  International Conference on}, pages 1--2. IEEE, 2014.

\bibitem{harp}
Prabhat~K. Gupta.
\newblock Xeon+fpga platform for the data center.
\newblock
  \url{http://www.ece.cmu.edu/~calcm/carl/lib/exe/fetch.php?media=carl15-gupta.pdf},
  2015.

\bibitem{sirius}
Johann Hauswald, Michael~A. Laurenzano, Yunqi Zhang, Cheng Li, Austin Rovinski,
  Arjun Khurana, Ronald~G. Dreslinski, Trevor Mudge, Vinicius Petrucci, Lingjia
  Tang, and Jason Mars.
\newblock Sirius: An open end-to-end voice and vision personal assistant and
  its implications for future warehouse scale computers.
\newblock In {\em Proceedings of the Twentieth International Conference on
  Architectural Support for Programming Languages and Operating Systems},
  ASPLOS '15, pages 223--238, New York, NY, USA, 2015. ACM.

\bibitem{parakeet}
Eric Hielscher.
\newblock {\em Locality Optimization For Data Parallel Programs}.
\newblock PhD thesis, New York University, 2013.

\bibitem{spartan}
Chien-Chin Huang, Qi~Chen, Zhaoguo Wang, Russell Power, Jorge Ortiz, Jinyang
  Li, and Zhen Xiao.
\newblock Spartan: A distributed array framework with smart tiling.
\newblock In {\em 2015 USENIX Annual Technical Conference (USENIX ATC 15)},
  pages 1--15, Santa Clara, CA, July 2015. USENIX Association.

\bibitem{hwkmeans}
H.M. Hussain, K.~Benkrid, H.~Seker, and A.T. Erdogan.
\newblock Fpga implementation of k-means algorithm for bioinformatics
  application: An accelerated approach to clustering microarray data.
\newblock In {\em Adaptive Hardware and Systems (AHS), 2011 NASA/ESA Conference
  on}, pages 248--255, June 2011.

\bibitem{micro14lee}
HyoukJoong Lee, Kevin~J. Brown, Arvind~K. Sujeeth, Tiark Rompf, and Kunle
  Olukotun.
\newblock Locality-aware mapping of nested parallel patterns on gpus.
\newblock In {\em Proceedings of the 47th Annual IEEE/ACM International
  Symposium on Microarchitecture}, IEEE Micro, 2014.

\bibitem{maxeler}
{Maxeler Technologies}.
\newblock {MaxCompiler white paper}, 2011.

\bibitem{mencer2011finding}
Oskar Mencer, Erik Vynckier, James Spooner, Stephen Girdlestone, and Oliver
  Charlesworth.
\newblock Finding the right level of abstraction for minimizing operational
  expenditure.
\newblock In {\em Proceedings of the fourth workshop on High performance
  computational finance}, pages 13--18. ACM, 2011.

\bibitem{scala}
M.~Odersky.
\newblock Scala.
\newblock \url{http://www.scala-lang.org}, 2011.

\bibitem{baidu}
Jian Ouyang, Shiding Lin, Wei Qi, Yong Wang, Bo~Yu, and Song Jiang.
\newblock Sda: Software-defined accelerator for largescale dnn systems.
\newblock Hot Chips 26, 2014.

\bibitem{catapultdnn}
Kalin Ovtcharov, Olatunji Ruwase, Joo-Young Kim, Jeremy Fowers, Karin Strauss,
  and Eric~S. Chung.
\newblock Accelerating deep convolutional neural networks using specialized
  hardware.
\newblock Technical report, Microsoft Research, February 2015.

\bibitem{haskell}
Simon {Peyton Jones}~[editor], John Hughes~[editor], Lennart Augustsson, Dave
  Barton, Brian Boutel, Warren Burton, Simon Fraser, Joseph Fasel, Kevin
  Hammond, Ralf Hinze, Paul Hudak, Thomas Johnsson, Mark Jones, John
  Launchbury, Erik Meijer, John Peterson, Alastair Reid, Colin Runciman, and
  Philip Wadler.
\newblock {Haskell}~98 --- {A} non-strict, purely functional language.
\newblock Available from \url{http://www.haskell.org/definition/}, feb 1999.

\bibitem{pouchet10phd}
Louis-No{\"e}l Pouchet.
\newblock {\em Interative Optimization in the Polyhedral Model}.
\newblock PhD thesis, University of Paris-Sud 11, Orsay, France, January 2010.

\bibitem{pouchet11popl}
Louis-No{\"e}l Pouchet, Uday Bondhugula, C{\'e}dric Bastoul, Albert Cohen,
  J.~Ramanujam, P.~Sadayappan, and Nicolas Vasilache.
\newblock Loop transformations: Convexity, pruning and optimization.
\newblock In {\em 38th ACM SIGACT-SIGPLAN Symposium on Principles of
  Programming Languages (POPL'11)}, pages 549--562, Austin, TX, January 2011.
  ACM Press.

\bibitem{pouchet13fpga}
Louis-Noel Pouchet, Peng Zhang, P.~Sadayappan, and Jason Cong.
\newblock Polyhedral-based data reuse optimization for configurable computing.
\newblock In {\em Proceedings of the ACM/SIGDA International Symposium on Field
  Programmable Gate Arrays}, FPGA '13, pages 29--38, New York, NY, USA, 2013.
  ACM.

\bibitem{catapult}
Andrew Putnam, Adrian~M. Caulfield, Eric~S. Chung, Derek Chiou, Kypros
  Constantinides, John Demme, Hadi Esmaeilzadeh, Jeremy Fowers, Gopi~Prashanth
  Gopal, Jan Gray, Michael Haselman, Scott Hauck, Stephen Heil, Amir Hormati,
  Joo-Young Kim, Sitaram Lanka, James Larus, Eric Peterson, Simon Pope, Aaron
  Smith, Jason Thong, Phillip~Yi Xiao, and Doug Burger.
\newblock A reconfigurable fabric for accelerating large-scale datacenter
  services.
\newblock In {\em Proceeding of the 41st Annual International Symposium on
  Computer Architecuture}, ISCA '14, pages 13--24, Piscataway, NJ, USA, 2014.
  IEEE Press.

\bibitem{chimps}
Andrew~R. Putnam, Dave Bennett, Eric Dellinger, Jeff Mason, and Prasanna
  Sundararajan.
\newblock Chimps: A high-level compilation flow for hybrid cpu-fpga
  architectures.
\newblock In {\em Proceedings of the 16th International ACM/SIGDA Symposium on
  Field Programmable Gate Arrays}, FPGA '08, pages 261--261, New York, NY, USA,
  2008. ACM.

\bibitem{pldi13halide}
Jonathan Ragan-Kelley, Connelly Barnes, Andrew Adams, Sylvain Paris, Fr{\'e}do
  Durand, and Saman Amarasinghe.
\newblock Halide: A language and compiler for optimizing parallelism, locality,
  and recomputation in image processing pipelines.
\newblock In {\em Proceedings of the 34th ACM SIGPLAN Conference on Programming
  Language Design and Implementation}, PLDI '13, pages 519--530, New York, NY,
  USA, 2013. ACM.

\bibitem{rompf12optimizing}
Tiark Rompf, Arvind~K. Sujeeth, Nada Amin, Kevin Brown, Vojin Jovanovic,
  HyoukJoong Lee, Manohar Jonnalagedda, Kunle Olukotun, and Martin Odersky.
\newblock Optimizing data structures in high-level programs.
\newblock POPL, 2013.

\bibitem{kiwi}
Satnam Singh and David~J. Greaves.
\newblock Kiwi: Synthesis of fpga circuits from parallel programs.
\newblock In {\em Proceedings of the 2008 16th International Symposium on
  Field-Programmable Custom Computing Machines}, FCCM '08, pages 3--12,
  Washington, DC, USA, 2008. IEEE Computer Society.

\bibitem{smith2005scientific}
M.C. Smith, Jeffrey~S Vetter, and Sadaf~R. Alam.
\newblock Scientific computing beyond {CPU}s: {FPGA} implementations of common
  scientific kernels.
\newblock In {\em Proceedings of the 8th Annual Military and Aerospace
  Programmable Logic Devices International Conference}, 2005.

\bibitem{delite-tecs14}
Arvind~K. Sujeeth, Kevin~J. Brown, HyoukJoong Lee, Tiark Rompf, Hassan Chafi,
  Martin Odersky, and Kunle Olukotun.
\newblock Delite: A compiler architecture for performance-oriented embedded
  domain-specific languages.
\newblock In {\em TECS'14: ACM Transactions on Embedded Computing Systems},
  July 2014.

\bibitem{ecoop13sujeeth}
Arvind~K. Sujeeth, Tiark Rompf, Kevin~J. Brown, HyoukJoong Lee, Hassan Chafi,
  Victoria Popic, Michael Wu, Aleksander Prokopec, Vojin Jovanovic, Martin
  Odersky, and Kunle Olukotun.
\newblock Composition and reuse with compiled domain-specific languages.
\newblock In {\em European Conference on Object Oriented Programming}, ECOOP,
  2013.

\bibitem{opencl}
{The Khronos Group}.
\newblock {OpenCL 2.0}.
\newblock \url{http://www.khronos.org/opencl/}.

\bibitem{venkat}
Anand Venkat, Mary Hall, and Michelle Strout.
\newblock Loop and data transformations for sparse matrix code.
\newblock In {\em Proceedings of the 36th ACM SIGPLAN Conference on Programming
  Language Design and Implementation}, PLDI 2015, pages 521--532, New York, NY,
  USA, 2015. ACM.

\bibitem{zaharia10spark}
Matei Zaharia, Mosharaf Chowdhury, Michael~J. Franklin, Scott Shenker, and Ion
  Stoica.
\newblock Spark: cluster computing with working sets.
\newblock In {\em Proceedings of the 2nd USENIX conference on Hot topics in
  cloud computing}, HotCloud'10, pages 10--10, Berkeley, CA, USA, 2010. USENIX
  Association.

\bibitem{zhang2005reconfigurable}
GL~Zhang, Philip Heng~Wai Leong, Chun~Hok Ho, Kuen~Hung Tsoi, Chris~CC Cheung,
  Dong-U Lee, Ray~CC Cheung, and Wayne Luk.
\newblock Reconfigurable acceleration for monte carlo based financial
  simulation.
\newblock In {\em Field-Programmable Technology, 2005. Proceedings. 2005 IEEE
  International Conference on}, pages 215--222. IEEE, 2005.

\bibitem{autopilot}
Zhiru Zhang, Yiping Fan, Wei Jiang, Guoling Han, Changqi Yang, and Jason Cong.
\newblock Autopilot: A platform-based esl synthesis system.
\newblock In Philippe Coussy and Adam Morawiec, editors, {\em High-Level
  Synthesis}, pages 99--112. Springer Netherlands, 2008.

\bibitem{zhuo2008high}
Ling Zhuo and Viktor~K Prasanna.
\newblock High-performance designs for linear algebra operations on
  reconfigurable hardware.
\newblock {\em Computers, IEEE Transactions on}, 57(8):1057--1071, 2008.

\end{thebibliography}

\end{document}